# Ranking the locations and predicting future crime occurrence by retrieving news from different Bangla online newspapers

**[1]Md. Jumman Hossain, [2]Rajib Chandra Das , [3]Md. Ruhul Amin, [4]Md. Saiful Islam**

**[1] Department of Computer Science and Engineering, Shahjalal University of Science and Technology, Sylhet, Bangladesh**
*jummanhossain92@gmail.com*

**[2] Department of Computer Science and Engineering, Shahjalal University of Science and Technology, Sylhet, Bangladesh**
*aporba.das@gmail.com*

**[3] Department of Computer Science and Engineering,Shahjalal University of Science and Technology Sylhet, Bangladesh**
*ruhul92@live.com*

**[4] Department of Computer Science and Engineering,Shahjalal University of Science and Technology Sylhet, Bangladesh**
*saiful-cse@sust.edu*

## Abstract

There have thousands of crimes are happening daily all around. But people keep statistics only few of them, therefore crime rates are increasing day by day. The reason behind can be less concern or less statistics of previous crimes. It's much more important to observe the previous crime statistics for general people to make their outing decision and police for catching the criminals are taking steps to restrain the crimes and tourists to make their travelling decision.

National institute of justice releases crime survey data for the country, but does not offer crime statistics up to Union or Thana level. Considering all of these cases we have come up with an approach which can give an approximation to people about the safety of a specific location with crime ranking of different areas locating the crimes on a map including a future crime occurrence prediction mechanism. Our approach relies on different online Bangla newspapers for crawling the crime data, stemming and keyword extraction, location finding algorithm, cosine similarity, naïve Bayes classifier and a custom crime prediction model.

## 1. Introduction

Most crime rankings are based on crimes per thousand residents which immediately creates an unfair playing field if you get thousands of tourists or workers per day. Those thousands of "outsiders" will inevitably commit crimes or inadvertently create opportunities for crime that would not exist in cities or states not getting a lot of tourists or daily workers. Even people feel worried about taking decisions for outing; where should they go and where don't.

Our goal is to design a future crime occurring prediction system alongside area based crime ranking for basically focusing on Bangladesh. Where, to get the crime occurrence of different locations have been collected after crawling different news from different Bangla online newspaper. This system also provides a map where different crime has been pointed according to their occurrence zone. This will definitely be helpful for general people, tourists to decide their outing and police to know where the crime is happening frequently.

## 2. Related Work

Keivan Kianmehr and Reda Alhajj presented a support vector machine (SVM) based approach to predict the location as alternative to existing modeling approaches.

SVM forms the new generation of machine learning techniques used to find optimal separability between classes within datasets. [1]

Xifan Zheng, Yang Cao and Zhiyu Ma introduced a model to predict the future serial crime location with aid of criminal geographic profiling and the time and locations of past crime scenes. In this model they have taken the effect of distance decay and the local geographic features into consideration and formulate a probability density function of the future serial crime site to realize the prediction. [2]

B. Chandra , Manish Gupta and M. P. Gupta introduced a novel approach based on dynamic time wrapping and parametric Minkowski model to find similar crime trends among various crime sequences of different crime locations and subsequently use this information for future crime trends prediction.[3]

We have understood from our studies that most of the system used existing crime databases and did not provide up to date solution. As our approach relies on crime data from daily newspapers, so this will surely provide a better solution for ranking crime occurrence and future crime prediction.

## 3. News Crawling

To work with crime data, the first thing we required is – crawling the news from different online newspapers and parse the required info. For performing the crawling, we have used a pre-developed crawler from our research lab. The crawled news contains –the title of the news, domain name, published date and the location is some cases. From this crawled we have parsed news parsed title, content, category, city, domain, date, URL and type.

## 4. Indexing Root Words and Top Words

To attain a better performance it's often helpful to remove commonly used words or stop words such as "করা", "হয়" etc. This process of removing stop words from text is called stop listing. The stemming process normalizes words by conflating a number of morphologically similar words to a single root form or stem. For example, "খেলার", "খেলায়", "খেলতে" all are reduced to "খেলা"।

Every document has a list of words containing useless words and useful words. we have found out a set of stop words: {"এক", "নামের", "করে", "তিনি", "রয়েছেন", "জানান", "তার", "করে", "দিয়ে", "থাকা", "নিয়ে", "যায়", "ছিল", "বলে", "আছে", "প্রথমে", "যাওয়া", "হয়", "পরে", "করা", "শুনেছি", "কিনা", "ভাল", "দেখা", "হচ্ছে"}. And the rest words are

Index>
<filePath>F:\Data WareHouse\small repository\Crawler_Data\www.amadershomoy2.com\12-7-2012-12-48-16</filePath>
<byteInfo>1 3700861728 3700861791 3700861791 3700870240</byteInfo>
<indexed>true</indexed>
<TITLE> রাজধানীতে ছিনাতাইকারীর ছুরিকাঘাতে নিহত ১, আহত ১</TITLE>
<CONTENT> রাজধানীতে ছিনাতাইকারীর ছুরিকাঘাতে নিহত ১, আহত ১ || অসধফবৎৎঝয়ড়সঢ়. ঈড়স ( আমাদের সময়. কম) জুলাই ১২, ২০১২, বৃহস্পতিবার : আষাঢ় ২৮, ১৪১৯ । আপডেট বাংলাদেশ সময় রাত ১২:০০ অৎপঘরাব আজজ৫�৬বকর পাতাসমঞ্ছৎ রাজধানীতে ছিনাতাইকারীর ছুরিকাঘাতে নিহত ১, আহত ১ নিজস্ব প্রতিবেদক আমাদের সময়. কম ইসমাইল হোসেন ইমু ও জোনায়েদ মানসুর : রাজধানীর মহাখালীতে ছিনাতাইকারীর ছুরির আঘাতে মাছ ব্যবসায়ী নিহত. তার নাম হাফিজ উদ্দিন ( ৪০) । এ ঘটনায়ে ইউসুফ ( ১৫) নামে আরেক জন আহত হয়েছেন। আজ ভোর সাড়ে পাচটার দিকে এ ঘটনা ঘটে। ইউসুফকে ঢাকা মেডিকেল কলেজ হাসপাতালে ভর্তি করা হয়েছে। ইউসুফ হাসপাতালে সাংবাদিকদের জানায়, তারা দু" জনই গাজীপুর এলাকার মাছ ব্যবসায়ী। এ ব্যাপারে শিল্পাঞ্চল থানায় পুলিশ আজ বেলা সাড়ে ১১টা পর্যন্ত এ ঘটনা জানেনা বলে আমাদের সময়. কমকে জানায়। বিস্তারিত আসছে- - - - - - - - - স্থানীয় সময়: ১২. ১১ ঘণ্টা, ১২ জুলাই ২০১২ বদরুল বোরহান / </CONTENT>
<CATEGORY> অন্যান্য</CATEGORY>
<CITY> ঢাকা গাজীপুর</CITY>
<DOMAIN> www.amadershomoy2.com</DOMAIN>
<DATE> 20120712062</DATE>
<URL> http://www.amadershomoy2.com/content/2012/07/12/middle0103.htm</URL>
<TYPE> news</TYPE>
<PATH> F:\Data WareHouse\small repository\Crawler_Data\www.amadershomoy2.com\12-7-2012-12-48-16</PATH>
<BYTE_INFO> 1 3700861728 3700861791 3700861791 3700870240</BYTE_INFO>
</Index>

Fig. 1  Sample crawled data format .

considered as main word (Keyword). We have studied more than 500 news articles and stored more than 1000 stop words in a text file.

ঢাকার বনানী এলাকায় এক বিকাশ প্রতিনিধিকে কুপিয়ে ৭ লাখ টাকা ছিনতাই করেছে দুর্বৃত্তরা। রোববার দুপুর দেড়টায় রাজধানীর সবুজবাগ থানার বাসাবো ওয়াসা রোডে জিন ইন্টারন্যাশনাল নামের বিকাশ এজেন্সির বিক্রয় প্রতিনিধি এনামুল হককে (৪০) কুপিয়ে সাত লাখ টাকা প্রতিনিধি করে। তিনি ঢাকা মেডিকেল কলেজ হাসপাতালে ( ঢামেক) চিকিৎসাধীন রয়েছেন। এনামুল হক জানান, ওই স্থানে ৫/৬ দুর্বৃত্ত তার গতিরোধ করে ধারালো অস্ত্র দিয়ে কুপিয়ে হাতে থাকা টাকার ব্যাগটি নিয়ে যায়। ব্যাগে সাত থেকে আট লাখ টাকা ছিল বলে তিনি জানান। ঢামেকের কর্তব্যরত চিকিৎসক জানান, এনামুল হকের হাতে, বুকে ও পিঠে ধারালো অস্ত্রের আঘাত আছে। জিন ইন্টারন্যাশনালের সুপারভাইজার তানভির নেওয়াজ খান জানান, খবর পেয়ে প্রথমে এনামুল হককে বাসাবো জেনারেল হাসপাতালে নিয়ে যাওয়া হয়। পরে ভাল চিকিৎসার জন্য তাকে ঢামেকে স্থানান্তর করা হয়। সবুজবাগ থানার ভারপ্রাপ্ত কর্মকর্তা ( ওসি) বাবুল মিয়া এ প্রসঙ্গে জানান, মারামারির ঘটনা শুনেছি। ছিনতাই কীনা খতিয়ে দেখা হচ্ছে।

Fig. 2  Document, containing useful and useless words.

Storing root words is not so easier like stop word. There are more than 40,000 words are used in Bangladeshi news articles. In order to compute root word from a sample word we have to study about bangla grammartical rules (পদ, প্রকৃতি, প্রত্যয়, উপসর্গ, কারক, বিভক্তি, সন্ধি-বিচ্ছেদ) and also required some natural language processing. But that was not our concern. So we have stored more than 4,500 words and corresponding root words in a text file.

পুলিশে পুলিশ
পুলিশকে পুলিশ
পুলিশের পুলিশ
পুলিশও পুলিশ
পুলিশদের পুলিশ
পুলিশরা পুলিশ
খেলায় খেলা
খেলার খেলা
খেলতে খেলা
খেলাকে খেলা

Fig. 3  Sample word to root word map.

## 5. Extracting Top Words From News

From our studies it has been observed that every Pronoun (সর্বনাম) and Conjunction (অব্যয়) are stop words. Pronouns and Conjunctions are not used to define a news category. Pronouns and Conjunction are added to stop word list. We can also add several verb (ক্রিয়া) and Adjective (বিশেষণ) words to stop word list. Actually we need all the Nouns (বিশেষ্য), several verbs (ক্রিয়া) and adjectives (বিশেষণ).

We have picked top words from about 250 news articles and counted term frequency for every term. The term which is not a stop word and exists more than 5 times counted as a top word. In later, the term which is more frequent in news article, picked as a top word.

পুলিশ →crime = 200 →sports = 0 →entertainment = 0 →technology = 0  →others = 9 , Total: 209

দল →crime = 24 →sports = 61 →entertainment = 1 →technology = 1 →others = 31 ,Total: 118

ডাকাত →crime = 72 →sports = 0 →entertainment = 0 →technology = 0  →others = 0 , Total: 72

মামলা →crime = 55 →sports = 0 →entertainment = 2 →technology = 2 →others = 2 , Total: 61

শিশু →crime = 32 →sports = 0 →entertainment = 2 →technology = 0  →others = 25 , Total: 59

খুন →crime = 55 →sports = 0 →entertainment = 0  - →technology = 0 →others = 0, Total: 55

Fig. 4  Term frequency of different words.

## 6. Categorizing the News

To categorize the news articles Naïve Bayes text classification approach has been followed considering performance compared to other approaches. The probability of a document d being in class c is computed as

Here,

$$P(c) = \frac{Number\ of\ document\ in\ category, c}{Number\ of\ total\ document}$$

nd = Number of term in Document d

$$P(t_k|c) = \frac{Term\ Frequency\ in\ category,\ c}{Term\ Frequency\ in\ all\ documents}$$

$$\hat{P}(t|c) = \frac{T_{ct}}{\sum_{t' \in V} T_{ct'}},$$

Now we will compute probability for every category individually. Maximum value will be defined by specific category.

$$c_{map} = \arg\max_{c \in \mathbb{C}} \hat{P}(c|d) = \arg\max_{c \in \mathbb{C}} \hat{P}(c) \prod_{1 \le k \le n_d} \hat{P}(t_k|c).$$

In this equation many conditional probabilities are multiplied, one for each position $1 \le k \le n_d$. This can result in a floating point underflow. It is therefore better to perform the computation by adding logarithms of probabilities instead of multiplying probabilities. The class with the highest log probability score is still the most probable. So the equation will be changed to,

$$c_{map} = \arg\max_{c \in \mathbb{C}} \left[\log \hat{P}(c) + \sum_{1 \le k \le n_d} \log \hat{P}(t_k|c)\right].$$

To eliminate zeros, we use add-one technique, which simply adds one to each count.

$$\hat{P}(t|c) = \frac{T_{ct} + 1}{\sum_{t' \in V}(T_{ct'} + 1)} = \frac{T_{ct} + 1}{(\sum_{t' \in V} T_{ct'}) + B},$$

Now, Data for parameter estimation example for Naïve Bayes Clustering:

If any new news articles has been arrived then their categorization result calculates like below:

Sample document:

ডাকাতি মামলার আসামিকে কুপিয়ে হত্যা।ঢাকার কেরানীগঞ্জে বাবুল ওরফে হক বাবুল ( ৩২) নামের এক ব্যক্তিকে কুপিয়ে হত্যা করেছে দুর্বৃত্তরা।

Calculated probabilities using Naïve Bayes clustering algorithm.

Crime: -0.47712125471966244
Sports: -3.112605001534575
Entertainment: -3.112605001534575
Technology: -3.4136349971985562
Others: -3.4136349971985562

Category Calculated: crime (As it is the Maximum Probability)

| Serial No. | News Categorization | |
|---|---|---|
| | Contents | *category* |
| 01 | ১৪ বছর আগে চট্টগ্রামের বহদ্দারহাটে ছাত্রলীগের গাড়িতে হামলা চালিয়ে আট জনকে হত্যার যে মৃত্যুদণ্ডে দণ্ডিত চার আসামির সবাই আপিলের রায়ে খালাস পেয়েছেন। | Crime |
| 02 | টি- টোয়েন্টি বিশ্বকাপের চ্যাম্পিয়ন শ্রীলঙ্কা এখনো পর্যন্ত টি- টোয়েন্টি বিশ্বকাপের সবগুলো ম্যাচেই জয়লাভ করেছে ভারত। আইসিসি টি২০ বিশ্বকাপের ফাইনালে ভারতকে ৬ উইকেটে হারিয়ে শিরোপা নিজেদের ঘরে নিয়েছে শ্রীলঙ্কা। | Sports |
| 03 | টিভি পর্দায় জাহিদ হাসানের নানা রূপ। পুরনো রূপ ভেঙে তিনি আবার নতুন রূপে হাজির হয়েছেন ' নজিরবিহীন নজির আলী' নাটকে। লিখেছেন মাহবুব হাসান জ্যোতি ' আমি ভাই ভিন্ন ধরনের চরিত্রের কাঙাল - আলাপচারিতার গুরুত্বেই বললেন জাহিদ হাসান। আরমান ভাইয়ের চরিত্র ছাপিয়ে তিনি এখন নজর আলী হয়ে উঠেছেন। | Entertainment |
| 04 | ভোরবেলা যাচ্ছিলাম কমলাপুর স্টেশনে। রাস্তায় খুব কম যানবাহন। তবু এক সিগন্যালে একটু থামতে হলো বাঁ দিক থেকে কয়েকটি গাড়ি ক্রসিং পার হয়ে ডান দিকে আসায়। একটি পিকআপ পাশ ঘেঁষে থামতে বাধ্য হলো। অত ভোরে কোনো বোকা চালকও লালবাতি মানেন না। পিকআপের দুটো চটের বস্তা। একটি বস্তার মুখের দিকে সামান্য ফাঁক দিয়ে দেখা গেল একটি মরা মুরগির পা ও পাখনা। চল্লিশ-পঞ্চাশ বছর আগে হলে মনে করতাম মরা মুরগি কুড়িয়ে কেউ ডাস্টবিনে ফেলে দিতে নিয়ে যাচ্ছে। এখন মনে পড়ল অন্য কথা। আমার কর্তব্য ছিল গাড়িটিকে আটকে পুলিশকে খবর দেওয়া। তা না করতে পারায় গ্লানি ও অপরাধ বোধ করি। | Others |
| 05 | গ্যালাক্সি এস৫ স্মার্টফোনটির একটি মিনি বা ছোট সংস্করণ আসছে। গ্যালাক্সি এস৫ এর এ সংস্করণটি হবে পানি- রোধী। স্যামসাং নিউজিল্যান্ডের অফিশিয়াল ওয়েবসাইটে এ তথ্য জানানো হয়েছে। অবশ্য মিনি সংস্করণটির তথ্য এখনও আনুষ্ঠানিকভাবে ঘোষণা করেনি দক্ষিণ কোরিয়ার প্রতিষ্ঠানটি। | Technology |

Fig. 5 Categorized news from example data.

## 7. Extracting Location and Date

For ranking the locations based on the crime scene it is obvious to find the exact crime locations. To do this first we store the locations. Here, we map all the Thana and district's name and store them on the database. It'll be more accurate if we store the corresponding union name. We find the location at the time of finding the root word and stop word. Here is the location finding algorithm.

```
FindLocation(Doc)
    A = set of all stored locations;
    V = ∅;
    L = ∅;
    Foreach word ∈ Doc
        word = Findroot(word);
        If word ∈ A
            Loc = getLoc(word);
            V = {V ∪ Loc};
            Continue;
    Foreach location ∈ V
        If location.type = "thana"
            L = {L, locatio0n}
```

| News Id | Thana | District |
|---|---|---|
| Document 1 | দোহার | ঢাকা |
| Document 2 | রাজনগর | মৌলভীবাজার |

Fig. 5  Result on applying location finding algorithm.

cosine similarity. Measurement of cosine similarity ensures the similarity between different news. Similarity helps to distinguish the similar type crime in the particular area. We also use this cosine similarity to find the same news published in different newspapers. For finding the same news first we calculate the cosine similarity but cosine similarity is not enough for this. We also find the occurrence date and the location where this crime scene happened. If the date and location is same and similarity value is greater than a threshold value then we decide that this documents are same. Suppose we have several news having words like "ছিনতাই", "খুন", "ডাকাতি" and here we have calculated the TF value of those words in the below documents.

---

1.
দোহার উপজেলায় প্রবাসী নুরুল ইসলাম মাঝির বাড়িতে ডাকাতি হয়েছে। গত শনিবার রাতে উপজেলার নারিশা ইউনিয়নের ঝনকি গ্রামে এ ডাকাতি হয়। রাত ২টার দিকে ১২/১৫ জনের একটি সংঘবদ্ধ ডাকাত দল দেশীয় অস্ত্র নিয়ে উপজেলার নারিশা ইউনিয়নের ঝনকি গ্রামের কুয়েত প্রবাসী নুরুল ইসলাম মাঝির বাড়িতে হানা দেয়। ডাকাত দল বসত বাড়ির মূল গেটের কাঠের দরজা ভেঙে ভেতরে প্রবেশ করে। সেসময়ে পরিবারের সকলকে অস্ত্রের মুখে জিম্মি করে হাত বেঁধে আলমারিতে থাকা নগদ ২ লাখ ৫০ হাজার টাকা, ১০ ভরি স্বর্ণলংকার, ২টি মোবাইল সেট লুটে নেয়।

2.
রাজধানী রামপুরার বনশ্রীতে দুর্ধর্ষ ডাকাতির ঘটনা ঘটেছে। মঙ্গলবার দিবাগত গভীর রাতে বনশ্রীর এফ ব্লকে ৪ নম্বর রোডের ২৩ নম্বর বাসায় এ ঘটনা ঘটে। জানা যায়, গভীর রাতে ৫/৭ জনের ডাকাত দল ওই বাসার কলাপসিবল গেট খুলে ভেতরে প্রবেশ করে। দুই দারোয়ানকে হাত পাকিং করা একটি পালসার ব্র্যান্ডের মোটরসাইকেল নিয়ে যায় ও একটি প্রাইভেটকারের (ঢাকা মেট্রো-গ-৩৫-৩৫৯৯) যন্ত্রাংশ খুলে নিয়ে যায়।

---

Similarity between these two documents is 73.2799%

## 9. Finding and Removing the Same News

A single news can be published in different newspapers. For calculating the exact crime occurrence, we have to remove the repetitive news from the sample data. To calculate this we use cosine similarity for matching the document and then find the location and published date. If

---

1.
দোহার উপজেলায় প্রবাসী নুরুল ইসলাম মাঝির বাড়িতে ডাকাতি হয়েছে। গত শনিবার রাতে উপজেলার নারিশা ইউনিয়নের ঝনকি গ্রামে এ ডাকাতি হয়। রাত ২টার দিকে ১২/১৫ জনের একটি সংঘবদ্ধ ডাকাত দল দেশীয় অস্ত্র নিয়ে উপজেলার নারিশা ইউনিয়নের ঝনকি গ্রামের কুয়েত প্রবাসী নুরুল ইসলাম মাঝির বাড়িতে হানা দেয়। ডাকাত দল বসত বাড়ির মূল গেটের কাঠের দরজা ভেঙে ভেতরে প্রবেশ করে। সেসময়ে পরিবারের সকলকে অস্ত্রের মুখে জিম্মি করে হাত বেঁধে আলমারিতে থাকা নগদ ২ লাখ ৫০ হাজার টাকা, ১০ ভরি স্বর্ণলংকার, ২টি মোবাইল সেট লুটে নেয়।

2.
রাজনগরে ডাকাতি, আহত ৩ রাজনগর (মৌলভীবাজার) প্রতিনিধি | ৩০ মার্চ ২০১৪, রবিবার, ৯:৩৬ রাজনগরে একদিনের ব্যবধানে আবারও দুর্ধর্ষ ডাকাতির ঘটনা ঘটেছে। ৮-১০ জনের ডাকাতদল অস্ত্রের মুখে জিম্মি করে ১৩ ভরি স্বর্ণালংকার, নগদ ১ লাখ ৫০ হাজার টাকাসহ বিভিন্ন মালামাল লুট করে নিয়ে যায়। ডাকাতদের হামলায় মহিলাসহ ৩ জন আহত হয়েছেন। আহতরা হলেন রন্টু পাল (৫০) তার স্ত্রী গীতা রানী পাল (৪০) ও ছেলে রনি পাল (১৫)। আহতদের বিভিন্ন মৌলভীবাজার ২৫০ শয্যার হাসপাতালে ভর্তি করা হয়েছে। শুক্রবার গভীর রাতে দক্ষিণ টেংরা গ্রামের রন্টু পালের বাড়িতে মুখোশ পড়া ৮- ১০ জনের একদল ডাকাত হানা দেয়।

Fig. 6  Example data to apply location finding algorithm.

## 8. Finding Similarity of Different News

To find the similarity between different news, first we have calculated the Term Frequency (TF) of the inputted news, then the Inverse Document Frequency (IDF) and finally the the crime occurrence date and locations are same and similarity is greater than a threshold value then we can assume that this two documents are same. Suppose we have N documents. Now doc$_i$ and doc$_j$ two same news from different news source. If the similarity between this documents S$_{(i, j)}$ is greater than 60% and the location of loc$_i$ and loc$_j$ is same and there publishing date date$_i$ and date$_j$ is same then this two document is similar. If there are N documents then the complexity is O(n$^2$). Because we need

1.

পাঞ্জাবকে দিয়ে জয়খরা কাটাল মুম্বাই, স্পোর্টস ডেস্ক, বাংলানিউজটোয়েন্টিফোর.কম

মুম্বাই: ওয়াংখেড়েতে ফিরে আইপিএলের সপ্তম আসরে ষষ্ঠ ম্যাচে এসে জয়ের দেখা পেল মুম্বাই ইন্ডিয়ান্স। গত আসরের নিজেদের মাঠে অজেয় দলটি হারাল এবারের টুর্নামেন্টে প্রথম পাঁচটিতেই জেতা কিংস ইলেভেন পাঞ্জাবকে।

কিংস ইলেভেন পাঞ্জাব: ১৬৮/৫ (২০ ওভার)

মুম্বাই ইন্ডিয়ান্স: ১৭০/৫ (১৯.১ ওভার)

ফল: মুম্বাই জয়ী পাঁচ উইকেটে।

পাঞ্জাবকে ছুড়ে দেওয়া ১৬৯ রানের লক্ষ্যে নেমে শুরুতে চোখ ধাঁধানো কোনো ইনিংস খেলেনি গতবারের চ্যাম্পিয়নরা। শেষ তিন ওভারে তাদের প্রয়োজন ছিল ৪১ রান। কিন্তু কাইরন পোলার্ড ও আদিত্য তারের শেষ সময়ের ঝড়ে পাঁচ বল বাকি থাকতে জয় পেল তারা।

পোলার্ড ১১ বলে দুটি করে চার ও ছয়ে ২৮ রানে অপরাজিত ছিলেন। তারে খেলেছিলেন ছয় বলে একটি করে চার ও ছয়ে ১৬ রানের হার না মানা ইনিংস। এর আগে ২৩ রানের মধ্যে দুই উইকেট হারালেও চিদাম্বরম গৌতম ও অধিনায়ক রোহিত শর্মার ব্যাটে এগিয়ে যায় মুম্বাই। চিদাম্বরম ২৯ বলে ৩৩ ও রোহিত ৩৪ বলে চারটি চার ও দুটি ছয়ে ৩৯ রান করেন। এটাই সেরা ইনিংস। এছাড়া কোরি এন্ডারসন ৩৫ রানের দ্বিতীয় সেরা ব্যাটিং করেন। ২৫ বলে তিনটি চার ও দুটি ছয়ে সাজানো ইনিংস খেলে ম্যাচসেরা হয়েছেন নিউজিল্যান্ডের এই তারকা।

সন্দীপ শর্মা ও রিশি ধাওয়ান পাঞ্জাবের পক্ষে দুটি করে উইকেট নেন।

এর আগে টস জিতে ব্যাট করতে নেমে ২৪ রানের মধ্যে দুই উইকেট হারিয়ে বিপদে পড়েছিল পাঞ্জাব। তবে রিদ্ধিমান সাহা ও গ্লেন ম্যাক্সওয়েলের ব্যাটে লড়াই করার মতো সংগ্রহ করে দলটি। রিদ্ধিমান ৪৭ বলে চারটি চার ও তিনটি ছয়ে অপরাজিত ছিলেন ৫৯ রানে। ম্যাক্সওয়েল ২৭ বলে পাঁচটি বাউন্ডারি ও দুটি ওভার বাউন্ডারিতে ৪৫ রানে আউট হন।

to calculate all pair similarity.

Fig. 7 Document 1 for to similarity finding

The similarity is: 92.30530291471328
Location Found in Doc1: null;
Location Found in Doc2: null;
Date published: 03-05-2014

Document 1 Source :

Document 2 Source :

2.

পোলার্ড মালিঙ্গায় মুম্বাইয়ের প্রথম জয়

স্পোর্টস ডেস্ক, বিডিনিউজ টোয়েন্টিফোর ডটকম

নিজেদের মাঠে খেলতে নেমেই হারের গণ্ডি থেকে বেরিয়ে এল মুম্বাই ইন্ডিয়ান্স। কাইরন পোলার্ডের শেষমুহূর্তের ঝড়ো ব্যাটিংয়ে এবারের আসরের সবচেয়ে সফল দল কিংস ইলেভেন পাঞ্জাবকে ৫ উইকেটে হারিয়েছে তারা। তবে পাঞ্জাব ইনিংসের শেষ দিকে অসাধারণ বল করা লাসিথ মালিঙ্গার অবদানও কম নয়।

ছয় ম্যাচে গতবারের চ্যাম্পিয়ন মুম্বাইয়ের এটা প্রথম জয়। সমান সংখ্যক ম্যাচে পাঞ্জাবের এটা প্রথম হার। এই হারে রান রেটে পিছিয়ে পড়ে চেন্নাই সুপার কিংসের কাছে শীর্ষ স্থান হারিয়েছে তারা।

শনিবার পাঞ্জাবের ৫ উইকেটে গড়া ১৬৮ রানের লক্ষ্য ৫ বল হাতে রেখেই অতিক্রম করে যায় মুম্বাই। শেষ ৩ ওভারে জয়ের জন্য স্বাগতিক দলের প্রয়োজন ছিল ৪১ রানের। ১২ বলে 'ক্যামিও' ইনিংসে দুটি করে ছক্কা ও চার মেরে অনায়াসে দলকে জয়ের বন্দরে পৌঁছে দেন ক্যারিবীয় অলরাউন্ডার পোলার্ড। ২৮ রানে অপরাজিত ছিলেন তিনি।

মুম্বাইয়ের ওয়াংখেড়ে স্টেডিয়ামে লক্ষ্য তাড়া করতে নেমে ২৩ রানের মধ্যে ২ উইকেট হারিয়ে শুরুতেই বিপদে পড়ে গিয়েছিল স্বাগতিকরা। তবে উইকেটরক্ষক চিদাম্বরম গৌতম ও অধিনায়ক রোহিত শর্মার ৪১ বলে ৪৭ রানের জুটিতে সে ধাক্কা সামলে ওঠে তারা। ২৯ বলে ৩৩ রান করেন গৌতম। তারপরও রোহিত শর্মা ও কোরি অ্যান্ডারসনের ব্যাটে ভর করে জয়ের পথেই ছিল তারা। কিন্তু পরপর দুই ওভারে তাদের বিদায়ে আবারও শঙ্কায় পড়ে যায় দলটা।

রোহিত করেন ৩৪ বলে ৩৯ রান আর অ্যান্ডারসনের ব্যাট থেকে আসে ২৫ বলে ৩৫ রান।

তবে পোলার্ড আর আদিত্য তারের ১৫ বলে অবিচ্ছিন্ন ৪৪ রানের ঝড়ো ম্যাচজয়ী জুটিতে সহজেই কাঙ্ক্ষিত জয় মেলে মুম্বাইয়ের। ৬ বলে ১টি করে চার ও ছক্কায় অপরাজিত ১৬ রান করেন তারে।

পাঞ্জাবের পক্ষে দুটি করে উইকেট নেন সন্দীপ শর্মা ও রিশি ধাওয়ান। এর আগে টস জিতে ব্যাট করতে নেমে ২৪ রানের মধ্যে দুই উদ্বোধনী ব্যাটসম্যানকে হারিয়ে ধাক্কা খায় পাঞ্জাব। তবে গ্লেন ম্যাক্সওয়েলের আক্রমণাত্মক ব্যাটে বিপদ কাটিয়ে ওঠে আসরের সবচেয়ে সফল দলটি। উইকেটরক্ষক ঋদ্ধিমান সাহার সঙ্গে ৫০ বলে ৬৯ রানের জুটি গড়েন প্রথম তিন ম্যাচে টানা তিনটি অর্ধশতক করা ম্যাক্সওয়েল। স্পিনার হরভজন সিংয়ের বলে বেন ডাঙ্কের হাতে ক্যাচ দিয়ে ফেরার আগে ৪৫ রান করেন তিনি। ২৭ বলের ইনিংসে ৫টি চার ও ২টি ছক্কা মারেন তিনি।

তবে ম্যাক্সওয়েল ফিরলেও ৫৯ রান করে অপরাজিত ছিলেন সাহা। তার ৪৭ বলের ইনিংসে ৪টি চার ও ৩টি ছক্কায় সাজানো। ৩৪ রান দিয়ে ২ উইকেট নেন মুম্বাইয়ের হরভজন সিং

Fig. 8 Document 2 for to similarity finding

Result:
Is source same: No
Result:
Is the Document Same: Yes

## 10. Mapping the Extracted Data

To rank the crime occurrence of different locations we have pointed the crime in a map. We have stored the list of all location of Bangladesh consisting Division, District and Thana. The co-ordinate of all location relative to this map is also stored in a table. Now , when we found a location of a crime then added dynamically a dot to the specific location.

When we found frequent location for different crime news, we just make it a larger dot than the previous with a limit width. We also changed the color of dot redder.

## 11. Measure Crime Occurence Probability

With our retrieved crime statistics consisting crime zones, crime types and incident times, we have designed a statistical approach based on previous crime statistics to predict future crime occurring probability.

We want to predict the crime occurring probability of a specific zone in a specific month.

Our crime predictions parameters are:

$C_{ZM}$ = Number of crimes in a specific zone in a specific month.

$C_{TM}$ = Number of crimes in all zones in a specific month.

$C_Z$ = Number of crimes in a specific zone in all months.

$C_T$ = Total number of crimes.

Let, we have a list of zone, Z = {ঢাকা, সিলেট, কুমিল্লা, চট্টগ্রাম,...,রংপুর}

List of month, M = {January, February,... ,December}

List of year, Y = {2000....2013}

Idx1 = index of specific month.

Idx2 = index of specific zone.

We can define,

$$C_{ZM} = \sum_{i=1}^{Yn} C(i, idx1, idx2)$$

$$C_{TM} = \sum_{i=1}^{Yn} \sum_{k=1}^{Zn} C(i, idx1, k)$$

$$C_Z = \sum_{i=1}^{Yn} \sum_{j=1}^{Mn} C(i, j, idx2)$$

$$C_T = \sum_{i=1}^{Yn} \sum_{j=1}^{Mn} \sum_{k=1}^{Zn} C(i, j, k)$$

So, Probability of Next Crime Occurrence in a Specific Zone and Specific Month

$P_{ZM}$ = ($C_{ZM}$ /$C_{TM}$)*($C_Z$ /$C_T$). When ($C_{ZM}$, $C_{TM}$, $C_Z$, $C_T$)>0

With this crime prediction mechanism we can roughly predict the next crime occurrence probability of a specific area in a specific time.

## 11. Limitations

- We were required about 30000 root words but we worked with only 5000, so there have an option improve the root word finding algorithm.

- Here we do not finding any keywords actually, to categorize the news we have taken only the specific top words. Better accuracy can be gained through finding the key words.

- To find the locations we have find only up to Thana but if there have any union that haven't identified yet.

- We have used a static map to show the crimes of specific locations, but it can be pointed better through using Google map.

## 12. Future Work

There have some important scope to develop our approach, like:

- Finding the keywords dynamically with designing a smart algorithm.

- Developing a dynamic algorithm for root word finding.

- We have tried to design a better crime prediction algorithm, but it's possible to design a better approach using more advanced machine learning techniques.

## 13. Conclusion

To reduce the number of crime occurrence it's required to predict the crime prone zone. Unfortunately there haven't any previous work been done to predict crime of a specific location with retrieving news from different Bangla online newspapers. Though, our developed approach is not giving perfect result but expecting this will surely be helpful for government, general people, police or tourists to decide their outing location.

## 14. Acknowledgement

We are grateful to our university research lab and Pipilika Bangla search engine for providing valuable data. We also thank to the authors and websites from which we get so much important knowledge and support for our paper.

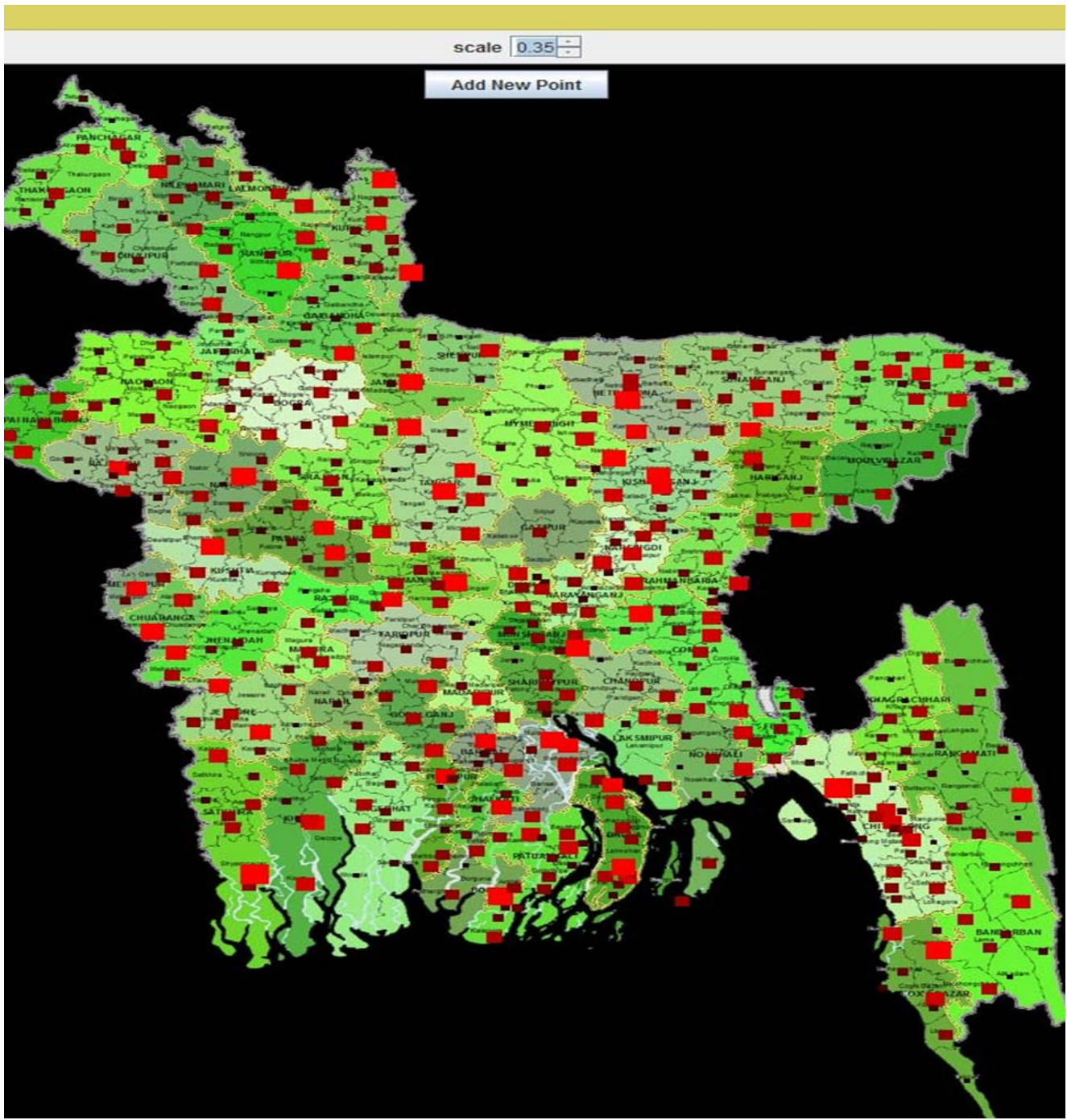

Fig. 9 Extracted crime data of different
locations pointed on Bangladesh Map